\documentstyle[psfig]{l-aa}
\def\la{\;
\raise0.3ex\hbox{$<$\kern-0.75em\raise-1.1ex\hbox{$\sim$}}\; }
\def\ga{\;
\raise0.3ex\hbox{$>$\kern-0.75em\raise-1.1ex\hbox{$\sim$}}\; }
\begin{document}
\thesaurus{11(02.12.1; 02.12.3; 11.17.1; 11.17.4 APM 08279+5255)}
\title{On the deuterium abundance at $z_{\rm a} = 3.514$
towards APM 08279+5255\thanks{The data presented herein were
obtained at the W. M. Keck Observatory, which is operated as a scientific
partnership among the California Institute of Technology, the University of
California and the National Aeronautics and Space Administration. The
Observatory was made possible by the generous financial support of the
W. M. Keck Foundation.}
}
\author{S. A. Levshakov\inst{1}
\thanks{On leave from the Ioffe 
Physico-Technical Institute, Russian Academy of Sciences, 
194021 St. Petersburg}
\and I. I. Agafonova\inst{1}$^{\;\, \star\star}$
\and W. H. Kegel\inst{2}}
\offprints{S.~A.~Levshakov}
\institute{
European Southern Observatory, 85748 Garching bei M\"unchen, Germany
\and
Institut f\"ur Theoretische Physik der Universit\"at Frankfurt am Main,
Postfach 11 19 32, 60054 Frankfurt/Main 11, Germany}
\date{Received November 00, 1999; accepted November 00, 1999}
\maketitle
\markboth{S.A. Levshakov et al.: D/H at $z_{\rm a} = 3.514$ towards
APM~08279+5255}{ }
%
\begin{abstract}
A very low primordial deuterium abundance of
${\rm D/H} \simeq 1.5\,10^{-5}$ has recently been proposed by
Molaro et al. in the Lyman limit system with 
$\log N_{{\rm H}\,{\sc i}} \simeq 18.1$ cm$^{-2}$
at $z_{\rm a} = 3.514$ towards the quasar APM~08279+5255. 
The D/H value was estimated through the standard 
Voigt fitting procedure utilizing a simple one-component model
of the absorbing region.
The authors assumed, however,
that `a more complex structure for the hydrogen cloud with somewhat
ad hoc components would allow a higher D/H'. 
We have investigated
this system using our new Monte Carlo inversion procedure 
which allows us to recover not only the physical parameters but also
the velocity and density distributions along the line of sight.
The absorption lines of H\,{\sc i}, C\,{\sc ii}, C\,{\sc iv},
Si\,{\sc iii}, and Si\,{\sc iv} were analyzed simultaneously.
The result obtained shows a 
considerably lower neutral hydrogen column density 
$\log N_{{\rm H}\,{\sc i}} \simeq 15.7$ cm$^{-2}$.
Hence, the measurement of the deuterium abundance in this system is
rather uncertain. 
We find that the asymmetric blue wing of the hydrogen Ly$\alpha$
absorption is readily explained by H\,{\sc i} alone.
Thus, up to now, deuterium was detected in only four QSO spectra
(Q~1937-1009, Q~1009+2956, Q~0130-4021, and Q~1718+4807) and all of them 
are in concordance with ${\rm D/H} \simeq  4\,10^{-5}$.

\keywords{line: formation -- line: profiles --
quasars: absorption lines -- 
quasars:individual: APM 08279+5255} 
\end{abstract}

\section{Introduction}

Accurate measurements of the hydrogen isotopic ratio D/H at high redshifts
may allow us to test experimentally the basis of the standard model of
big bang nucleosynthesis (BBN) -- its homogeneity. 
A homogeneous BBN implies
a {\it uniform} distribution of the D abundance
among the absorbing systems with low metallicity
since `no realistic astrophysical process other than the Big Bang could
produce significant D' (Schramm 1998, p.6).

First high quality spectral data of QSOs suggested, however, a dispersion in
D/H~$\equiv$ $N_{{\rm D}\,{\sc i}}/N_{{\rm H}\,{\sc i}}$
(the ratio of the D\,{\sc i} and H\,{\sc i} column densities)
of about one order of magnitude (for a summary, see Burles et al. 1999).
This finding provoked a lively discussion in the literature
on inhomogeneous models of BBN
(Dolgov \& Pagel 1999 and references cited therein).
But later, it was shown that 
a single D/H value of about $4\,10^{-5}$
is sufficient to describe 
all observations available up to now (Levshakov et al. 1998a, 1998b,
1999b).

\begin{figure}
\vspace{5.0cm}
\hspace{-2.2cm}\psfig{figure=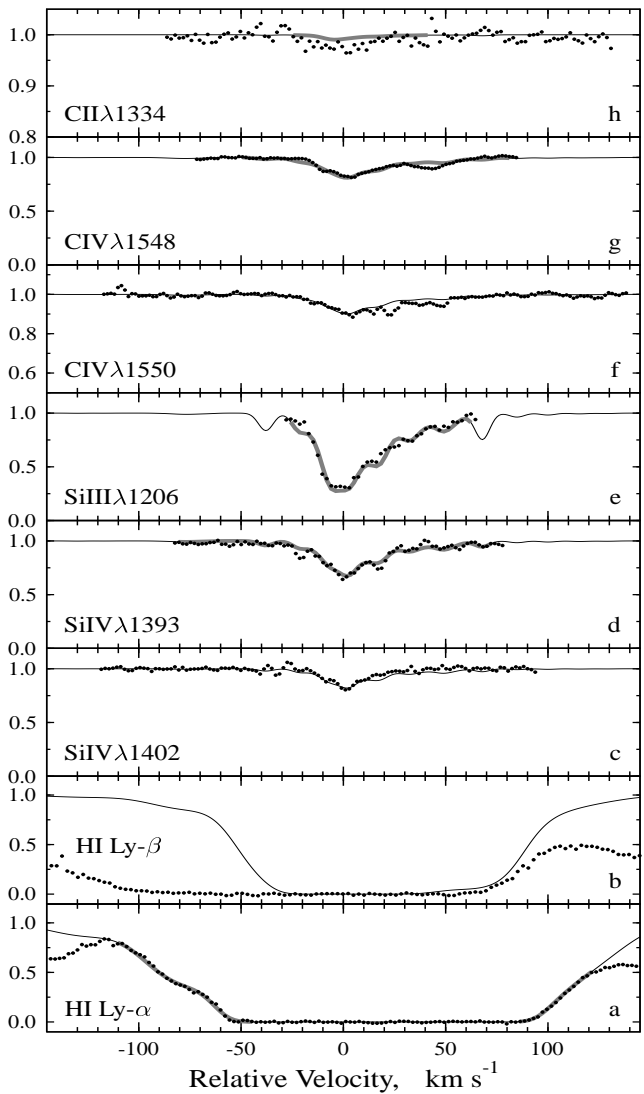,height=12.0cm,width=8.0cm}
\vspace{-2.5cm}
\caption[]{Observed and synthetic line profiles of the 
absorption system at $z_{\rm a} = 3.514$. 
Note different intensity scales in panels {\bf h} and {\bf f}. For details,
see text
}
\end{figure}

\begin{figure}
\vspace{6.0cm}
\hspace{-1.2cm}\psfig{figure=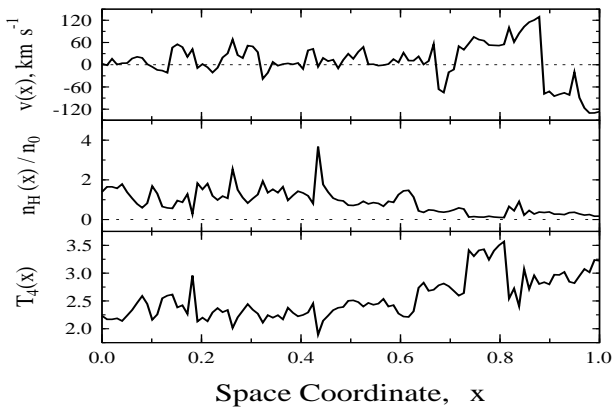,height=6.0cm,width=7.0cm}
\vspace{-6.5cm}
\caption[]{
MCI reconstruction of the radial velocity,
density, and kinetic temperature distributions. $T$ is given in units
of $10^4$~K
}
\end{figure}

A new result by Molaro et al. (1999, hereafter MBCV), 
if confirmed, could challenge again the uniformity
of the D/H space distribution because a very low deuterium abundance of
${\rm D/H} \simeq 1.5\,10^{-5}$ was found in an extremely low metallicity
system at $z_{\rm a} = 3.514$ towards APM 08279+5255.
MBCV consider the derived D abundance as a lower limit  
because their analysis was based on a simplified one-component model
of the absorbing cloud
which failed to fit the red wing of the Ly$\alpha$ line.
They note that the observed  
complex structure of C\,{\sc iv} and Si\,{\sc iv} implies the presence
of more than one component. 
They further
state that `additional components are required  
to reproduce the extra absorption on the red wing of Ly$\alpha$'
which would decrease the H\,{\sc i} column density for the major component
leading to a higher deuterium abundance. 

Following the MBCV suggestions, 
we have analyzed the absorption profiles of 
H\,{\sc i}, C\,{\sc ii}, C\,{\sc iv},
Si\,{\sc iii}, and Si\,{\sc iv} 
by using our new Monte Carlo inversion (MCI) algorithm (Levshakov et al.
1999c, hereafter LAK). 
In the present Letter we show that the actual neutral hydrogen column
density may be a factor of $\simeq 250$ lower than the value of MBCV, a fact
preventing any accurate measurements of the deuterium abundance in
the $z_{\rm a} = 3.514$ system.

The main difference between
MBCV's  analysis and our approach is that the former was 
performed in the framework of the microturbulent approximation 
using the standard Voigt fitting procedure,
whereas the latter
is based on a more general mesoturbulent model which describes the process
of line formation in the absorbing clouds more adequately
(Levshakov \& Kegel 1997).
In both cases the same spectra of APM 08279+5255 obtained with the Keck-I
telescope and the HIRES spectrograph by Ellison et al. (1999) are
utilized.

\section{Model assumptions and results}

Our model supposes a continuous absorbing gas slab of a thickness $L$
(presumably the outer region of a foreground galaxy). The absorber is assumed to
exhibit a mixture of bulk motions such as infall and outflows, tidal flows etc.
Then the gas motion along a given line of sight may be described by a fluctuating
(random) velocity field in which the velocities in neighboring volume
elements are correlated with each other (while in the standard microturbulent 
approximation completely uncorrelated velocities are assumed).
The gas is tenuous and optically thin in the Lyman continuum.
We are considering a compressible gas, i.e. the
total number density of hydrogen $n_{\rm H}$ is also a random function
of the space coordinate, $x$.
Following Donahue \& Shull (1991) and assuming that the ionizing radiation
field is constant, 
the ionization of different elements can be described by
one parameter only -- the ionization parameter 
$U \propto 1/n_{\rm H}$. Furthermore, 
for gas in thermal
equilibrium, Donahue \& Shull give an explicit relation between $U$
and the kinetic temperature $T$. 
The background ionizing spectrum in our model is taken in the form 
given by Mathews \& Ferland (1987). 

To estimate physical parameters and appropriate 
distributions of the velocity 
$v(x)$ and the normalized density
$y(x) = n_{\rm H}(x)/n_0$ , $n_0$ being the mean hydrogen density, 
we used the MCI procedure developed on the basis of
the reverse Monte Carlo technique (Levshakov et al. 1999a). 
In our computations, the continuous random functions $v(x)$ and $y(x)$
are represented by their sampled values at equally spaced intervals
$\Delta x$, i.e. by the vectors \{$v_1, ... , v_k$\} and
\{$y_1, ... , y_k$\} with $k$ large enough to represent
the narrowest components of the complex spectral lines.
For the ionization parameter as a function of $x$, we have
$U(x) = \hat{U}_0/y(x)$, with 
$\hat{U}_0$ being the reduced mean ionization parameter 
defined below.
We fix $z_{\rm a} = 3.51374$ (the value
adopted by MBCV) as a more or less arbitrary reference velocity
at which $v_j = 0$.

Our aim is to fit the model spectra simultaneously to the observed
hydrogen, carbon and silicon profiles. 
In this case the model requires the definition of a simulation
box for the six parameters (see LAK, for details)~:
the carbon and silicon abundances,
$Z_{\rm C}$ and $Z_{\rm Si}$, respectively, 
the rms velocity and density dispersions, 
$\sigma_{\rm v}$ and $\sigma_{\rm y}$,
respectively,
the reduced total hydrogen column density 
$\hat{N}_{\rm H} \left[= N_{\rm H}/(1+\sigma^2_{\rm y})^{1/2} \right]$, and 
the reduced mean ionization parameter 
$\hat{U}_0 \left[= U_0/(1+\sigma^2_{\rm y})^{1/2} \right]$. 
For the model parameters the following boundaries were adopted~:
$Z_{\rm C}$ ranges from $10^{-6}$ to $4\,10^{-4}$,
$Z_{\rm Si}$ from $10^{-6}$ to $3\,10^{-5}$,
$\sigma_{\rm v}$ from 25 to 80 km~s$^{-1}$, 
$\sigma_{\rm y}$ from 0.5 to 2.2,
$\hat{N}_{\rm H}$ from $5\,10^{17}$ to $8\,10^{19}$ cm$^{-2}$, and
$\hat{U}_0$ ranges from $5\,10^{-4}$ to $5\,10^{-2}$. 

Having specified the parameter space, we minimize the $\chi^2$ value.
The objective function includes the following portions 
of the absorption profiles  
(labeled by grey lines in Fig.~1)
which, 
after preliminary analysis, were chosen as most
appropriate to the MCI fitting~: for H\,{\sc i} Ly$\alpha$
$\Delta v$ ranges from $-110$ to $-47$ km~s$^{-1}$ (the blue wing)
and from 84 to 119 km~s$^{-1}$ (the red wing, see Fig.~1a),
for C\,{\sc ii} from $-26$ to 33  km~s$^{-1}$ (Fig.~1h),
for C\,{\sc iv}$\lambda1548$ from $-51$ to 79  km~s$^{-1}$ (Fig.~1g),
for Si\,{\sc iii} from $-28$ to 60  km~s$^{-1}$ (Fig.~1e), and
for Si\,{\sc iv}$\lambda1393$ 
$\Delta v$ ranges from $-82$ to 70  km~s$^{-1}$ (Fig.~1d). 
The necessity to choose these portions instead of the 
all available profiles was caused by a few discrepancies in the data.
Thus two small spikes in C\,{\sc iv}$\lambda1550$  
at $\Delta v = 15$ and 23 km~s$^{-1}$ are not seen in its stronger
blue counterpart C\,{\sc iv}$\lambda1548$, 
two minima in Si\,{\sc iv}$\lambda1393$ at $\Delta v = -21$ and 17  km~s$^{-1}$ 
are invisible in the profile of the red component 
Si\,{\sc iv}$\lambda1402$. But in Fig.~1 (panels {\bf f} and {\bf c})
the observed C\,{\sc iv}$\lambda1550$ and,
respectively, Si\,{\sc iv}$\lambda1402$
are shown together with the model spectra 
computed with the 
parameters derived from Ly$\alpha$, C\,{\sc ii}$\lambda1334$,
C\,{\sc iv}$\lambda1548$, Si\,{\sc iii}$\lambda1206$, and 
Si\,{\sc iv}$\lambda1393$ fitting to illustrate the consistency.
For the same reason the Ly$\beta$ 
model spectrum is shown in Fig.~1b 
at the expected position. 
The additional absorption on the blueward and the
redward sides of Ly$\beta$ in the observed spectrum
may be caused by Ly$\alpha$ and/or
Ly$\beta$ forest lines at different redshifts.
In Fig.~1, all model spectra are drawn by continuous thin lines,
whereas filled circles represent observations (normalized fluxes).

The MCI is a stochastic optimization procedure
and one does not know in advance 
if the global minimum of the objective function is reached in
a single run.
Therefore we executed several runs for the given data set
starting every calculation from a random point in the simulation box and
from completely random
configurations of the velocity and density fields.
The results for five runs are listed in Table~1. 
The first solution (upper row in Table~1) was used to calculate
the model spectra shown in Fig.~1, and
the corresponding distributions of $v(x)$, $y(x)$, and $T(x)$ which are 
presented in Fig.~2. 

\begin{table}
\centering
\caption{Cloud parameters derived from the Ly$\alpha$ and metal profiles
by the MCI procedure (the total hydrogen column density $N^{\rm H}_{18}$
in units of $10^{18}$ cm$^{-2}$, the neutral hydrogen column density
$N^{{\rm H}\,{\sc i}}_{15}$ in units of $10^{15}$ cm$^{-2}$,
the mean ionization parameter $U^0_{-2}$ in units of $10^{-2}$,
the rms velocity dispersion $\sigma_{\rm v}$ in  
km~s$^{-1}$,
the reduced $\frac{1}{\nu}\chi^2_{{\rm H}\,{\sc i}}$ 
per degree of freedom, $\nu = 42$, for the Ly$\alpha$
blue and red wings)
}
\label{tab1}
\begin{tabular}{rccccccc}
\hline
\noalign{\smallskip}
$N^{\rm H}_{18}$ & $N^{{\rm H}\,{\sc i}}_{15}$  &
  $U^0_{-2}$ & $\sigma_{\rm v}$ & $\sigma_{\rm y}$ &
  [C/H]$^\star$ & [Si/H]$^\star$ & 
  $\frac{1}{\nu}\chi^2_{{\rm H}\,{\sc i}}$ \\
\noalign{\smallskip}
\hline
\noalign{\smallskip}
 4.4 & 4.6 & 1.25 & 51 & 1.1 & $-1.7$ & $-0.6$ & 1.17\\
 5.0 & 5.3 & 1.10 & 50 & 0.8 & $-1.8$ & $-0.7$ & 1.15\\
 5.9 & 5.2 & 1.60 & 48 & 1.1 & $-1.8$ & $-0.7$ & 1.16\\
 8.7 & 7.2 & 2.54 & 70 & 1.6 & $-1.9$ & $-0.8$ & 1.16\\
 11.3 & 8.1& 1.83 & 52 & 0.9 & $-2.1$ & $-0.9$ & 1.03\\
\noalign{\smallskip}
\hline
\noalign{\smallskip}
\multicolumn{8}{l}{$^\star$\, based on solar abundances from Grevesse (1984);}\\
\multicolumn{8}{l}{
[X/H] = $\log (N_{\rm X}/N_{\rm H}) - \log (N_{\rm X}/N_{\rm H})_\odot$}
\end{tabular}
\end{table}

The median estimations of the parameters give
$N_{\rm H} = 5.9\,10^{18}$ cm$^{-2}$,
$N_{{\rm H}\,{\sc i}} = 5.3\,10^{15}$ cm$^{-2}$,
$U_0 = 1.6\,10^{-2}$, 
$\sigma_{\rm v} = 51$ km~s$^{-1}$,
$\sigma_{\rm y} = 1.1$,
[C/H] = -- 1.8, and [Si/H] = -- 0.7.
The results were obtained with $k = 100$ 
and the correlation coefficients 
$f_{\rm v} = f_{\rm y} = 0.95$. 

The MCI allowed us to fit precisely not only the blue wing of the
saturated Ly$\alpha$ line but the red one as well. 
In addition we have reached a reasonable
concordance between hydrogen and metal absorption lines. 
Although our model spectra of metal lines are consistent
with the observed profiles within the 3-$\sigma$ uncertainty
range, we found difficulties with the C\,{\sc ii} line fitting.
The C\,{\sc ii} model spectrum lies systematically over
the observed intensities (Fig.~1h). 
We note that the measurement of the C\,{\sc ii} line
depends sensitively on the exact definition of the continuum
because the line is very weak
(it was  
not included by Ellison et al. into  the $z_{\rm a} = 3.514$ system). 

Nevertheless, let us assume for a moment that
MBCV's identification of C\,{\sc ii} is correct, and let us 
further compare in some more detail 
the micro- and mesoturbulent results. Using the
Doppler parameters $b_{{\rm C}\,{\sc ii}} = 17.8$ km~s$^{-1}$ and
$b_{{\rm H}\,{\sc i}} = 21.0$ km~s$^{-1}$ measured by MBCV, it is
easy to estimate the turbulent velocity 
$\sigma_{\rm v, mic} \equiv b_{\rm turb}/\sqrt{2} = 12.4$ km~s$^{-1}$ 
and the kinetic temperature
$T_{\rm mic} = 8200$~K.
On the other side,
the MCI kinetic temperature distribution (Fig.~2)
does not reveal 
values lower than $2\,10^4$~K. 
Besides, $T_{\rm mic} = 8200$~K seems to be too
low for the photoionization heating.
The mean kinetic temperature from
the mesoturbulent solution $T_{\rm meso} = 25500$~K is,
probably, more realistic.
The four times higher mesoturbulent value of the rms velocity dispersion 
is also supported by observations.
Indeed, large broadening may be caused by the lensing effects
since this QSO exhibits two components of similar intensity separated by  
$\sim 0''.4$ (Irvin et al. 1998). 
The low $T_{\rm mic}$ and $\sigma_{\rm v, mic}$ 
seem to be an indication that the identification
of C\,{\sc ii} at $z_{\rm a} = 3.514$ is 
not very certain.

\section{Conclusions}

\begin{figure}
\vspace{6.3cm}
\hspace{-4.0cm}\psfig{figure=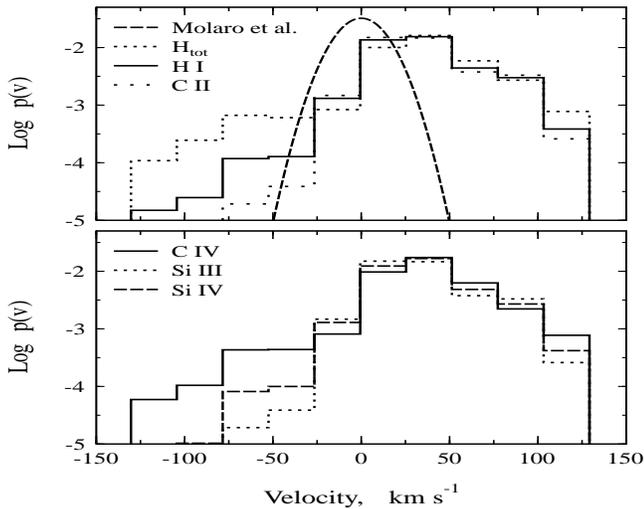,height=5.0cm,width=6.5cm}
\vspace{-4.5cm}
\caption[]{Density-weighted
radial velocity distribution functions, $p(v)$, for the total 
and neutral hydrogen, C\,{\sc ii}, C\,{\sc iv}, Si\,{\sc iii}, and
Si\,{\sc iv} as
restored by the MCI procedure. For comparison, in the upper panel
the short dashed line shows $p(v)$ for the homogeneous microturbulent model
adopted by MBCV
}
\end{figure}

We have shown that the determination of the D abundance in 
the $z_{\rm a} = 3.514$ 
system is strongly model dependent.
The Ly$\alpha$ profile can as well be modeled
with a considerably lower value of $N_{{\rm H}\,{\sc i}}$ as compared
to the MBCV model if one accounts for spacial correlations in the
large scale velocity and density fields. This implies that the
identification of the D\,{\sc i} line in this system 
cannot be confirmed or ruled out without additional 
observations of the higher order Lyman series lines in order to
constrain the total neutral hydrogen column
density and the velocity field configuration.

For this particular case, the presence of the extra-absorption
exactly at the expected deuterium position $\Delta v = -82$
km~s$^{-1}$ becomes evident from the
upper panel of Fig.~3 where dotted and solid line
histograms show, respectively, 
the density weighted radial velocity distributions
$p(v)$ for the total and neutral hydrogen, whereas the dashed
curve illustrates $p(v)$ as adopted by MBCV (a Gaussian
with $\sigma_{\rm v, mic} = 12.4$ km~s$^{-1}$). The radial velocity
distribution of H\,{\sc i} shows a blue-side asymmetry
which may just mimic the deuterium absorption.
The main maximum of $p_{{\rm H}\,{\sc i}}(v)$ is shifted to
$\Delta v \simeq 30$ km~s$^{-1}$ with respect to MBCV' frame of reference.
The distribution $p_{{\rm H}\,{\sc i}}(v)$ exhibits also an extended red wing
ranging up to $\Delta v \simeq 90$ km~s$^{-1}$.
This asymmetry is also clearly 
pronounced in the Si\,{\sc iii} profile (Fig.~1e).

The restored density
and velocity fields reveal a very complex structure 
which is manifested in  non-Gaussian distributions
as shown in Fig.~3 for the total hydrogen density as well as for the
individual ions. It appears that C\,{\sc iv} is a good tracer for
the distribution of the total hydrogen density. 

Since the Voigt profile fitting analysis may produce very misleading results
when applied to the case where the deviation from Gaussianity in
the velocity distribution becomes significant, 
the extremely low metallicity with 
${\rm [C/H]} \simeq -4.0$ and ${\rm [Si/H]} \simeq -3.5$ 
reported by MBCV
may be caused by their simplified model.
For example,
our analysis yields considerably higher values of
${\rm [C/H]} \simeq -1.8$ and ${\rm [Si/H]} \simeq -0.7$. 
A metallicity as high as [C/H] $\simeq -1$ and a silicon overabundance
[Si/C] $= 0.5 - 1$ dex has been measured in the metal systems at 
$z_{\rm a} \sim 4$ towards Q~0000--2619 (Savaglio et al. 1997).
A similar silicon overabundance has also been  observed in  
halo (population~II) stars.
The standard interpretation of these observations
is that the metal-poor gas was enriched by Type~II 
supernova nucleosynthesis products (e.g. Henry \& Worthey 1999).
Thus, our measurements are in agreement with these results.

Hence, it may be 
concluded that up to now deuterium lines have been identified
convincingly in only four QSO spectra.
Our previous D/H measurements in Q~1937--1009, Q~1009+2956, and
Q~1718+4807 and the last one by Kirkman et al. (1999) in Q0130--4021
are consistent with a single value of ${\rm D/H} \simeq 4\,10^{-5}$.

\begin{acknowledgements}
The authors are grateful to Ellison et al. for making their
data available. 
We thank Sandro D'Odorico for valuable comments
and Miguel Albrecht for kind advice in using the ESO computer cluster.
SAL and IIA gratefully acknowledge the hospitality of the 
European Southern Observatory (Garching), where this work
was performed.
\end{acknowledgements}


\begin{thebibliography}{}

\bibitem{}Burles, S., Kirkman, D., Tytler, D. 1999, ApJ, 519, 18 

\bibitem{}Dolgov, A. D., Pagel, B. E. J. 1999, New Astron., 4, 223

\bibitem{}Donahue, M., Shull, J. M. 1991, ApJ, 383, 511

\bibitem{}Ellison, S. L. et al. 1999, PASP, 111, 946

\bibitem{}Grevesse, N. 1984, Phys. Scripta, T8, 49

\bibitem{}Henry, R. B. C., Worthey, G. 1999, PASP, 111, 919

\bibitem{}Irwin, M. J., Ibata, R. A., Lewis, G. F., Totten, E. J. 1998, ApJ,
505, 529

\bibitem{}Kirkman, D., Tytler, D., Burles, S., Lubin, D., O'Meara, J. M.
1999, AAS, 194, 3001 (astro-ph/9907128)

\bibitem{}Levshakov, S. A., Kegel, W. H. 1997, MNRAS, 288, 787

\bibitem{}Levshakov, S. A., Kegel, W. H., Takahara, F. 1998a,
ApJ, 499, L1

\bibitem{}Levshakov, S. A., Kegel, W. H., Takahara, F. 1998b,
A\&A, 336, L29

\bibitem{}Levshakov, S. A., Kegel, W. H., Takahara, F. 1999a,
MNRAS, 302, 707 

\bibitem{}Levshakov, S. A., Tytler D., Burles S. 1999b,
in {\it Early Universe~: Cosmological Problems and Instrumental Technologies}
(Proceed. of the Gamov Memorial International Conference,
St.~Petersburg, August 23-28, 1999), in press (astro-ph/9812114)

\bibitem{}Levshakov, S. A., Agafonova, I. I., Kegel, W. H. 1999c, A\&A,
in preparation [LAK]

\bibitem{}Mathews, W. D., Ferland, G. 1987, ApJ, 323, 456

\bibitem{}Molaro, P., Bonifacio, P., Centurion, M., Vladilo, G. 1999,
A\&A, 349, L13 [MBCV]

\bibitem{}Savaglio, S., Cristiani, S., D'Odorico, S.,
Fontana, A., Giallongo, E., Molaro, P. 1997, A\&A, 318, 347

\bibitem{}Schramm, D. N. 1998, in {\it Primordial Nuclei and Their
Galactic Evolution}, eds. N. Prantzos, M. Tosi, \& R. von Steiger
(Kluwer Academic Publ.: Dordrecht, The Netherlands), 3


\end{thebibliography}
\end{document}